\batchmode
\makeatletter
\def\input@path{{\string"D:/Weberszpil/Calculo Fracional/Artigos 2013/Dirac-Pauli-G-factor/\string"/}}
\makeatother
\documentclass[english,aps,preprint]{paper}
\usepackage[LGR,T1]{fontenc}
\usepackage[latin9]{inputenc}
\usepackage{geometry}
\usepackage{textcomp}
\usepackage{amsmath}
\usepackage{amssymb}
\usepackage{setspace}
\usepackage{esint}
\makeatletter

\DeclareRobustCommand{\greektext}{%
  \fontencoding{LGR}\selectfont\def\encodingdefault{LGR}}
\DeclareRobustCommand{\textgreek}[1]{\leavevmode{\greektext #1}}
\DeclareFontEncoding{LGR}{}{}
\DeclareTextSymbol{\~}{LGR}{126}

\numberwithin{equation}{section}
\numberwithin{figure}{section}

\makeatother

\usepackage{babel}
\begin{document}

\title{Anomalous g-Factors for Charged Leptons in a Fractional Coarse-Grained
Approach}

\author{J. Weberszpil$^{1}$ J. A. Helayël-Neto$^{2}$}

\maketitle
$^{1}$josewebe@ufrrj.br

Universidade Federal Rural do Rio de Janeiro, UFRRJ-IM/DTL\\
 Av. Governador Roberto Silveira s/n- Nova Iguaçú, Rio de Janeiro,
Brasil

$^{2}$helayel@cbpf.br

Centro Brasileiro de Pesquisas Físicas-CBPF-Rua Dr Xavier Sigaud 150,\\
 22290-180, Rio de Janeiro RJ Brasil. 
\begin{abstract}
In this work, we investigate aspects of the electron, muon and tau
gyromagnetic ratios ($g-factor$) in a fractional coarse-grained scenario,
by adopting a Modified Riemann-Liouville (MRL) fractional calculus.
We point out the possibility of mapping the experimental values of
the specie's $g-factors$ into a theoretical parameter which accounts
for fractionality, without computing higher-order QED calculations.
We wish to understand whether the value of ($g-2$) may be traced
back to a fractionality of space-time.The justification for the difference
between the experimental and the theoretical value $g=2$ stemming
from the Dirac equation is given in the terms of the complexity of
the interactions of the charged leptons, considered as pseudo-particles
and ''dressed'' by the interactions and the medium. Stepwise, we build
up a fractional Dirac equation from the fractional Weyl equation that,
on the other hand, was formulated exclusively in terms of the helicity
operator. From the fractional angular momentum algebra, in a coarse-grained
scenario, we work out the eigenvalues of the spin operator. Based
on the standard electromagnetic current, as an analogy case, we write
down a fractional Lagrangian density, with the electromagnetic field
minimally coupled to the particular charged lepton. We then study
a fractional gauge-like invariance symmetry, formulate the covariant
fractional derivative and propose the spinor field transformation.
Finally, by taking the non-relativistic regime of the fractional Dirac
equation, the fractional Pauli equation is obtained and, from that,
an explicit expression for the fractional $g-factor$ comes out that
is compared with the experimental CODATA value. Our claim is that
the different lepton species must probe space-time by experiencing
different fractionalities, once the latter may be associated to the
dressing of the particles and, then, to the effective interactions
of the different families with the medium.

\vspace*{5mm}

\end{abstract}

\section{\noindent Introduction}

The electron spin magnetic moment, also called electron spin $g-factor$
$g_{e}$, is a dimensionless coupling parameter that appears in the
spin-orbit interaction and leads to the splitting in the atomic energy
levels, giving rises to the fine structure. Its value should not be
confused with the related gyromagnetic $\gamma$ ratio. The factor
$g_{e}$ appears in the Zeeman effect, as a coupling factor, in the
description of the interaction of atoms with external magnetic fields.

Some authors argue that the anomalous magnetic moment (AMM) of the
muon can be considered one of the most promising observables that
can give rises to a rich new physics, specially in particle physics
\cite{Passera 2008}. Other authors claim that the muons AMM is one
of the most precisely measured quantities in particle physics, reaching
a precision of 0.54ppm \cite{Fred Jegerlehner-The Muon g-2-2009}.
This argumentation is also reinforced because of the relatively large
value of the muons mass \cite{Eduardo de Rafael 2012}.

The tau anomaly is not very well known experimentally and, by this
reason do not provide a good test of the Standard Model (SM) \cite{Eduardo de Rafael 2012}.

The non-relativistic quantum mechanics predict by the reduction of
the Dirac equation to the Pauli equation that $g_{e}=2$. Experimentally,
the electron $g-factor$ has been measured with high precision, and
the ''2010 CODATA recommended values''\cite{CODATA} for this parameter
is $g_{e,exp}=2.00231930436153(53)$ that is supported by quantum
electrodynamics (QED) and is slightly different from that of Dirac's
theoretical prevision by quantum mechanics (QM). A good review on
the introduction and history of $g-factors$, up to 2007, can be found
in ref. \cite{Introduction and history of g-factors-Eduardo de Rafael}.
For a nonperturbative approach, the reader can see the ref. \cite{nonperturbative}.
An update of the Electron and Muon g-Factors can be found in ref.
\cite{Eduardo de Rafael 2012}.

In this article, we investigated the fractional coarse-grained aspects
of the electron anomalous magnetic $g-factor$, showing the possibility
of mapping the experimental $g-factor$ with some theoretical fractionality
parameter, so that we take the viewpoint that fractionality may be
responsible for the deviation from the $g-2$ quantum-mechanical result.
The possible justification for the experimental difference value from
the theoretical Dirac $g_{e}=2$ is given in the realm of complexity
of interactions for the electrons, considered as pseudo-particle ''dressed''
with the interactions and the medium. Here, we look at the dynamical
system as an open system that can interact with the environment and
we argue that fractional calculus can be an important tool to study
open classical and quantum systems. \cite{Tarasov 2012-Fract Oscillator}\textbf{
.}

Fractional Calculus (FC) is one of the generalizations of the classical
calculus. It provides a redefinition of mathematical tools and it
seems very useful to deal with anomalous \cite{klafter,klafter2,klafter3,9,scalas,Hilfer2}
and frictional systems. Several applications of FC may be found in
the literature \cite{stanisvasky,Metzler,Glockle}. Presently, areas
such as field theory and gravitational models demand new conceptions
and approaches which might allow us to understand new systems and
could help in extending well-known results. Interesting problems may
be related to the quantization of field theories for which new approaches
have been proposed \cite{Cresus e Everton,Kleinert,baleanu,goldfain}.
In conection with to our work, it is worthy to mention here that a
fractional Riemann\textendash{}Liouville Zeeman effect, an atempt
to implement gauge invariance in fractional field theories and an
angular momentum algebra proposed with the Riemann-Liouvile formalism
here reported in the paper of Ref. \cite{Fract Field Herrmann}. Low-energy
nuclear excitations was studied by the constuction of a fractional
symmetric rigid rotor in order to study barionic excitations \cite{Rigid Rotor-Herrmann}.

Here, we claim that the use of an approach of FC based on a sequential
form of modified Riemann-Liouville (MRL) fractional calculus \cite{Jumarie1}
is more appropriate to describe the dynamics associated with field
theory and particle physics in the space of nondifferentiable solution
functions, or in the coarse-grained space-time.

It seems that a reasonable way to probe the classical framework of
physics is to highlight that, in the space of our real world, the
generic point is not infinitely small (or thin), it rather has a thickness.
In a coarse-grained space, a point is not infinitely thin, and here,
this feature is modeled by means of a space in which the generic differential
is not $dx$, but rather $(dx)^{\alpha}$, and likewise for the time
variable t. It is noteworthy to highlight the ideas in the interesting
work by Nottale \cite{Nottale}, where the notion of fractal space-time
is first introduced. Non-integer differentiability and randomness
\cite{Grigolini-Rocco-West-1999} are mutually related in their nature,
in such a way that studies on fractals on the one hand, and fractional
Brownian motion on the other hand, are often parallel in the ref.
\cite{Nottale}. A function continuous everywhere, but nowhere integer-differentiable,
necessarily exhibits random-like or pseudo-random features, in that
various samplings of these functions, on the same given interval,
will be different. This may explain the huge amount of literature
extending the theory of stochastic differential equations to describe
stochastic dynamics driven by fractional Brownian motion \cite{Jumarie1,Jumarie-Lagrang Fract,Jumarie2}.

Recently, we have also used FC to analyze the well-established canonical
quantization symplectic algorithm \cite{Sympletic}. The focus was
to construct a generalized extension of that method to treat a broader
number of mechanical systems with respect to the standard method.
In this sense, we adopted the MRL prescription for fractional derivative.
There we argued that there is a number of problems in considering
classical systems besides the ones that involve the quantization of
second-class systems. These problems encompass the so-called nonconservative
systems. The peculiarity about them is that the great majority of
actual classical systems is nonconservative but, in spite of that,
the most advanced formalisms of classical mechanics deal only with
conservative systems \cite{RW}. Following \cite{Sympletic}, dissipation,
for example, is present even at the microscopic level. There is dissipation
in every non-equilibrium or fluctuating process, including dissipative
tunneling \cite{Cal} and electromagnetic cavity radiation \cite{Sen},
for instance. There we argued that one way to suitably treat nonconservative
systems is through Fractional Calculus, FC, since it can be shown
that, for example, a friction force has its form stemming from a Lagrangian
that contains a term proportional to the fractional derivative, which
may be a derivative of any non-integer order \cite{RW}.

Field theory aspects of non-linear dynamics are today an important
subject of study in different physical and mathematical sub-areas,
but the real success and a radically new understanding of non-linear
processes has occurred over the past 40 years. This understanding
was inspired by the discovery and insight of chaotic dynamics, where
the randomness of some physical processes are considered; more precisely,
when particle trajectories are indistinguishable for random process
\cite{GZ}.

In a previous work, also based on this approach, we have worked out
a suggested version of a fractional Schrödinger equation, with a lowest-order
relativistic correction, obtained from a fractional wave equation
\cite{Cresus-Jose-Helayel-caos} to which a mass term has been adjoined,
to give us a fractional Klein-Gordon equation (FKGE). With the definition
of some fractional operators, the McLaurin expansion and an ansatz
for the plane wave solutions, we have obtained fractional versions
of Bohmian equations to describe the particle dynamics associated
with Bohmian mechanics, in the space of non-integer differentiable
functions. We had also done a formulation for an anomalous dispersion
relation and to a refraction index, related to massless particle in
a coarse-grained media and a vacuum refractive index for a coarse-grained
non-trivial optical medium.

In Ref. \cite{Cresus-Jose-Helayel-caos},\textbf{ }we have argued
that the modeling of TeV-physics may demand an approach based on fractal
operators and FC and we claimed that, in the realm of complexity,
nonlocal theories and memory effects were connected to complexity
and also that the FC and the nondifferentiable nature of the microscopic
dynamics may be connected with time scales. Using the MRL definition
of fractional derivatives, we have worked out explicit solutions to
a fractional wave equation with suitable initial conditions to carefully
understand the time evolution of classical fields with fractional
dynamics. First, by considering space-time partial fractional derivatives
of the same order in time and space, a generalized fractional D'Alembertian
is introduced. By means of a transformation of variables to light-cone
coordinates, an explicit analytical solution was obtained. Also, aspects
connected with Lorentz symmetry were analyzed with two different approaches.

Here, to achieve our goals, we carefully build up a fractional Dirac
equation from fractional Weyl equation that, on the other hand, was
constructed with helicity and projection concepts. We begin by constructing
the fractional angular momentum algebra in a coarse-grained scenario
in order to justify the fractional spin operator eigenvalue, used
to project the spin in the linear momentum direction. Also, a covariant
fractional derivative is obtained by proposing a spinor field transformation
that leads to a fractional potential vector field gauge transformation.
Then, the minimal coupling to the electromagnetic field is naturally
obtained by writing down a fractional Lagrangian density with an included
term that takes into account the electronic current that, in turn,
is based on the electromagnetic current. So, we are not proposing
the matter transformation from the very start \cite{Fract Field Herrmann};
we rather get it from the matter-gauge coupling together with the
gauge field transformation. Finally, by an non-relativistic approximation
of the fractional Dirac equation, the fractional Pauli equation is
obtained that leads to an explicit expression for the fractional g-factor.
We point out that fractional Dirac equation has already been studied
by several authors over the past decade \cite{Plyushchay-Cubic root of Klein-Gordon,Raspini,Zavada}.

Our paper is outlined as follows: In Section 2, we consider the mathematical
background, with some expressions of the fractional coarse-grained
calculus and the modified Riemann-Liouville fractional derivative.
Section 3 contains the Fractional Angular Momentum Algebra, in a coarse-grained
scenario. Section 4 is devoted to Fractional Field Equations: Weyl
and Dirac and the results for g-factor for electron, muon and tau
are presented. In Section 5 we present the Gordon Decomposition for
Fractional Dirac Equation and also the Fractional Spin Current. Finally,
in Section 6, we cast our Discussion and Conclusions.

\section{Mathematical Background}

In the sequence we use the an alternative approach by considering
fractional coarse-grained space-time instead of fractional space functions,
meaning that nor the space nor the time are infinitely thine but have
''thickness''.

As the use of certain calculation rules are essential to our approach,
we briefly comment on this point, before presenting these rules. It
seems to us worthy to note that the chain rule, as well as the product
rule of Leibniz, for the fractional approach used here, had their
validity mathematically proven in view of the recent demonstration
of the formal fractional Taylor expansion, in the context of the modified
Riemann-Liouville (MRL) formalism \cite{Demonst Taylor}. We also
point out to the fact that to the date prior to the publication of
this statement, in its countersigned mathematical formal approach,
doubts about the validity of the rules could reasonably exist, because
of previous statements somewhat incomplete. However, given the existence
of this new theoretical formal support, published in an international
journal physica A, endorses the validity of the fractional Taylor
expansion and therefore proves it. The fractional Taylor expansion
is the mathematical basis for the validation of the chain rule and
also of the Leibniz product rule, in the MRL context. We also emphasize
that the rules so obtained can then be viewed as good approximations.
Than, we point out that the Leibniz rule used here is a good approximation
that comes from the first two terms of the fractional Taylor series
development, that holds only for nondifferentiable functions \cite{Jumarie- Acta Sin,Jumarie 2013}
and are as good and approximated as the classical integer one. These
approaches are quite similar to their counterparts made in the integer
order calculus and therefore so good as.

The well-tested definitions for fractional derivatives, so called
Riemann-Liouville and Caputo have been frequently used for several
applications in scientific periodic journals. In spite of its usefulness
they have some dangerous pitfalls. For this reason, recently it was
proposed an interesting definition for fractional derivative \cite{Jumarie1},
so called modified Riemann-Liouville (MRL) fractional derivative,
which is less restrictive than other definitions and its basic definition
is
\begin{eqnarray}
D^{\alpha}f(x) & = & {\displaystyle \lim_{x\rightarrow0}\, h^{-\alpha}\sum_{k=0}^{\infty}{\alpha \choose k}\, f\left(x+(\alpha\,-\, k)h\right)}=\nonumber \\
 & = & \frac{{1}}{\Gamma(1-\alpha)}\frac{d}{{dx}}\,\int_{0}^{x}(x\,-\, t)^{-\alpha}\left(f(t)-f(0)\right)dt;\\
 & 0<\alpha<1.\nonumber 
\end{eqnarray}
Some advantages can be cited, first of all, using the MRL definition
we found that derivative of constant is zero, and second, we can use
it so much for differentiable as non differentiable functions. They
are cast as follows \cite{Jumarie 2013}:

\subsubsection*{(i) Simple rules:}

\begin{align}
D^{\alpha}K & =0,\qquad Dx^{\gamma}=\frac{\Gamma(\gamma+1)}{\Gamma(\gamma+1-\alpha)}x^{\gamma-\alpha},\;\gamma>0,\\
(u(x)v(x))^{(\alpha)} & =u^{(\alpha)}(x)v(x)+u(x)v^{(\alpha)}(x).\label{eq:Leibniz}
\end{align}

\paragraph*{(ii) Simple Chain Rules:}

\begin{equation}
\frac{d^{\alpha}}{dx^{\alpha}}f[u(x)]=\frac{d^{\alpha}f}{du^{\alpha}}\,\left(\frac{du}{dx}\right)^{\alpha}.\label{eq:Chainrule nondif func-1}
\end{equation}
where $f$ is \textgreek{a}-differentiable and $u$ is differentiable
with respect to $x$ and, 

\begin{equation}
\frac{d^{\alpha}}{dx^{\alpha}}f[u(x)]=\frac{df}{du}\,\frac{d^{\alpha}u}{dx^{\alpha}}.\label{eq:chain rule space-time coarse-1}
\end{equation}
for coarse-grained space-time. Where $f(u(x))$is not differentiable
w.r.t $x$ but it is differentiable w.r.t $u$, and $u$ is not differentiable
w.r.t $x$.

For further details, the readers can follow the refs. \cite{Livro Jumarie,Jumarie- Acta Sin,Jumarie 2013}
which contain all the basic for the formulation of a fractional differential
geometry in coarse-grained space, and refers to an extensive use of
coarse-grained phenomenon.

Here, another comment is pertinent: the fractional MRL approach for
nondifferentiable functions has similar rules and has definition with
a mathematical limit operation comparable to certain definitions of
local fractional derivatives, as that introduced by Kolwankar and
Gandal \cite{Kolwankar1,Kolwankar 2,Kolwankar 3} with some studies
in the literature. For example, the works of Refs. \cite{calculus of local fractional derivatives,On the local fractional derivative,Carpintieri}
or the approaches with Hausdorff derivative, also called fractal derivative
\cite{Hausdorff or fractal derivative,comparative-Fractal -Fractional},
that can be applied to power-law phenomena and the recently developed
$\alpha-derivative$ \cite{Kobelev}. The MRL approach seems to us
to be an integral version of the calculus mentioned above and all
of them deserve to be more deeply investigated, under a mathematical
point of view, in order to give exact differences and similarities
respect to the traditional fractional calculus with Riemann-Liouville
or Caputo definition and with local fractional calculus and even fractional
q-calculus \cite{Richard Herrmann,Metzler,Caceres 2004}, as well
as in the comparative point of view of physics\cite{comparative-Fractal -Fractional,Caceres 2004,Bologna-Grigolini-1999},
for the scope of applicability.

Now that we have set up these fundamental expressions, we are ready
to carry out the calculations of main interest.

\section{Fractional Angular Momentum Algebra}

Here we will derive the commutation algebra for $spin-1/2$ particles
in a coarse-grained medium.

Since in this approach MRL the chain and Leibniz rules holds, it is
not difficult to obtain the commutation relation for position momentum
and defining the fractional angular momentum components \cite{Rigid Rotor-Herrmann,Fract Field Herrmann},
we can write

\begin{eqnarray}
\hat{L}_{z}^{\alpha}:= & \left[\hat{x}^{\alpha},\hat{p_{y}}^{\alpha}\right] & ,\\
\hat{L}_{y}^{\alpha}:= & \left[\hat{z}^{\alpha},\hat{p_{x}}^{\alpha}\right] & ,\\
\hat{L}_{x}^{\alpha}:= & \left[\hat{y}^{\alpha},\hat{p_{z}}^{\alpha}\right] & .
\end{eqnarray}

Again, one can obtain the commutation relations using the above definitions
and the chain\eqref{eq:Chainrule nondif func-1}, \eqref{eq:chain rule space-time coarse-1}
and Leibniz \eqref{eq:Leibniz} rules for MRL as

\begin{eqnarray}
\left[\hat{L}_{x}^{\alpha},\hat{L}_{y}^{\alpha}\right]= & i\Gamma(\alpha+1)\hbar^{\alpha}M_{x,\alpha}\hat{L}_{z}^{\alpha}\\
\left[\hat{L}_{z}^{\alpha},\hat{L}_{x}^{\alpha}\right]= & i\Gamma(\alpha+1)\hbar^{\alpha}M_{x,\alpha}\hat{L}_{y}^{\alpha}\\
\left[\hat{L}_{y}^{\alpha},\hat{L}_{z}^{\alpha}\right]= & i\Gamma(\alpha+1)\hbar^{\alpha}M_{x,\alpha}\hat{L}_{x}^{\alpha}
\end{eqnarray}

$\sigma_{\mu}=(\sigma_{0},\sigma_{1},\sigma_{2},\sigma_{3})=(I_{2},\sigma_{x},\sigma_{y},\sigma_{z})$

To build up the algebra, we define some the operators, as in the following.

The square of fractional angular moment operator is defined as in
ref. \cite{Rigid Rotor-Herrmann}: $\left(\hat{L}^{\alpha}\right)^{2}=\hat{L}^{\alpha}\hat{L}^{\alpha}=\left(\hat{L}_{x}^{\alpha}\right)^{2}+\left(\hat{L}_{y}^{\alpha}\right)^{2}+\left(\hat{L}_{z}^{\alpha}\right)^{2}$.
We also define the fractional angular momentum operator $\vec{L}^{\alpha}$and
the fractional raising and lowering operators, $L_{+}^{\alpha}$and
its hermitian conjugate $L_{-}^{\alpha},$ respectively as

$\vec{L}^{\alpha}\equiv\hat{L}_{x}^{\alpha}\hat{i}+\hat{L}_{y}^{\alpha}\hat{j}+\hat{L}_{z}^{\alpha}\hat{k}$
and $\begin{cases}
\hat{L}_{+}^{\alpha}\equiv & \hat{L}_{x}^{\alpha}+i\hat{L}_{y}^{\alpha}\\
L_{-}^{\alpha}\equiv & \hat{L}_{x}^{\alpha}-i\hat{L}_{y}^{\alpha}
\end{cases}$.

It follows, the commutation the relations

$\left[\hat{L}_{+}^{\alpha},\hat{L}_{-}^{\alpha}\right]=2\Gamma(\alpha+1)\hbar^{\alpha}M_{\alpha}\hat{L}_{z}^{\alpha},$

$\left[\hat{L}_{z}^{\alpha},\hat{L}_{+}^{\alpha}\right]=\Gamma(\alpha+1)\hbar^{\alpha}M_{\alpha}\hat{L}_{+}^{\alpha},$

$\left[\hat{L}_{z}^{\alpha},\hat{L}_{-}^{\alpha}\right]=-\Gamma(\alpha+1)\hbar^{\alpha}M_{\alpha}\hat{L}_{-}^{\alpha},$

$\left[\hat{L}^{\alpha}\hat{L}^{\alpha},\hat{L}_{+}^{\alpha}\right]=\left[\hat{L}^{\alpha}\hat{L}^{\alpha},\hat{L}_{-}^{\alpha}\right]=\left[\hat{L}^{\alpha}\hat{L}^{\alpha},\hat{L}_{x}^{\alpha}\right]=\left[\hat{L}^{\alpha}\hat{L}^{\alpha},\hat{L}_{y}^{\alpha}\right]=\left[\hat{L}^{\alpha}\hat{L}^{\alpha},\hat{L}_{z}^{\alpha}\right]=0.$

The above commutation relations indicates that the ordinary integer
angular momentum algebra do not change. This implies that the raising
and the lowering operators acting on a eigenstate $\left|j,m\right\rangle ,$
leads to an new state vector $\left|j,m\pm1\right\rangle $but with
eigenvalue $\Gamma(\alpha+1)\hbar^{\alpha}M_{\alpha},$ that is, it
raises the $\hat{L}_{z}^{\alpha}$eigenvalue by the latter increment.

Another important conclusion is that, for the ordinary Pauli spin
matrices, the representation for the basis of $\left(\hat{L}^{\alpha}\right)^{2}$is
the same as the integer case. This allows us to rewrite all usual
relations from ordinary quantum mechanics spin algebra\textbf{ }in
a coarse-grained scenario\textbf{. }The only difference is the instead
of $\hslash$ we have to substitute in all relations an effective
factor $\hslash_{eff}=\Gamma(\alpha+1)\hbar^{\alpha}M_{\alpha}.$
There is no fractional number of particles but, there exist an effective
Planck constant. 

We remark here that, if in the definition o fractional moment operator,
eq. \eqref{eq:Fractional Energy-Momenta Operators}, the complex $i$
were redefined as $i^{\alpha},$ the algebra probably could have completely
changed.

\section{Fractional Fields Equations: Weyl and Dirac}

The Weyl Equation is a relativistic wave equation for describing massless
$spin-1/2$ particles.

Remembering that the helicity is the projection of the spin onto the
direction of momentum, we proceed to pursue by this way to achievement
of the fractional Weyl equations in the following.

We write the projection of spin onto its linear momentum as

\begin{equation}
\lambda_{\alpha}=\overrightarrow{S^{\alpha}}\circ\frac{\overrightarrow{p^{\alpha}}}{\left|\overrightarrow{p^{\alpha}}\right|},\quad\lambda_{\alpha}=\pm1,\label{eq:projection of spin onto linear momentum}
\end{equation}
with the spin vector, $\overrightarrow{S_{\alpha}}$, obtained from
the angular momentum algebra, given by

\begin{equation}
\overrightarrow{S^{\alpha}}=\frac{\hbar^{\alpha}}{2}\Gamma(\alpha+1)M_{x,\alpha}\overrightarrow{\sigma}.
\end{equation}

Here the Pauli matrices are the same since the structure of the algebra
are not modified, as shown in section 3.

For a mass less particle, the relativistic energy-momentum expression
may be \cite{Aspects-nosso}

\begin{equation}
E^{\alpha}=\left|\overrightarrow{p^{\alpha}}\right|c^{\alpha}
\end{equation}

Considering a $2\times1$component spinorial field $\chi_{L,\alpha}$
belonging to a group representation . In term of this field and the
above equations, we can write

$\frac{\hbar^{\alpha}}{2}\Gamma(\alpha+1)M_{x,\alpha}\overrightarrow{\sigma}\circ\overrightarrow{p^{\alpha}}\chi_{L,\alpha}=\frac{\hbar^{\alpha}}{2}\Gamma(\alpha+1)M_{x,\alpha}\chi_{L,\alpha}$
or

\begin{equation}
c^{\alpha}\overrightarrow{\sigma}\circ\overrightarrow{p^{\alpha}}\chi_{L}=\left|\overrightarrow{p^{\alpha}}\right|\chi_{L,\alpha}=E^{\alpha}\chi_{L,\alpha}.\label{eq:First sigma-p-E-qui}
\end{equation}

In order to proceed with the adequate quantization, we have carry
on the correspondence principle. To this intention we propose the
fractional operators energy and momentum as

\begin{equation}
\begin{cases}
\widehat{E}^{\beta}=i\left(\hbar\right)^{\beta}\frac{\partial^{\beta}}{\partial t^{\beta}}\\
\widehat{p}^{\alpha}=-i\left(\hbar\right)^{\alpha}M_{x,\alpha}\frac{\partial^{\alpha}}{\partial x^{\alpha}}
\end{cases};\label{eq:Fractional Energy-Momenta Operators}
\end{equation}
where the constant $M_{x,\alpha}$ is included for dimensional reasons.

By correspondence principle, the eq. \eqref{eq:First sigma-p-E-qui}
can be rewritten as 
\begin{equation}
i\left(\hbar\right)^{\alpha}\overrightarrow{\sigma}\circ M_{x,\alpha}\frac{\partial^{\alpha}}{\partial x^{\alpha}}\chi_{L,\alpha}+i\left(\hbar\right)^{\alpha}\frac{M_{t,\alpha}}{c^{\alpha}}\frac{\partial^{\alpha}}{\partial t^{\alpha}}\chi_{L,\alpha}=0,
\end{equation}
 here we have considered the space and time with the same fractionality.

The above equation can be written in a co-variant form as a fractional
Weyl equation as 
\begin{equation}
i\left(\hbar\right)^{\alpha}\sigma^{\mu}\partial_{\mu}^{(\alpha)}\chi_{L,\alpha}=0.\label{eq:Fractional Weyl 1}
\end{equation}

Here $\sigma^{\mu}=(\sigma^{0},\sigma^{i})=(\sigma^{0},\sigma^{1},\sigma^{2},\sigma^{3})=(I_{2},\sigma_{x},\sigma_{y},\sigma_{z})$
are the usual Pauli-spin matrices and we defined the space-time fractional
derivative in the Minkowsky metric as $\partial_{\mu}^{(\alpha)}=(\frac{M_{t,\alpha}}{c^{\alpha}}\frac{\partial^{\alpha}}{\partial t^{\alpha}};M_{x,\alpha}\frac{\partial^{\alpha}}{\partial x^{\alpha}}).$

Also defining the conjugated Pauli-spin matrices as $\bar{\sigma}^{\mu}=(\sigma^{0},-\sigma^{i})$,
where the properties holds 
\begin{equation}
tr(\sigma^{\mu}\bar{\sigma}^{\nu})=tr(\bar{\sigma}^{\mu}\sigma^{\nu})=2\eta^{\mu\nu},
\end{equation}
or 
\begin{equation}
\sigma^{\mu}\bar{\sigma}^{\nu}+\sigma^{\nu}\bar{\sigma}^{\mu}=2\eta^{\mu\nu}\mathbf{1},
\end{equation}

\begin{equation}
\bar{\sigma}^{\mu}\sigma^{\nu}+\bar{\sigma}^{\nu}\sigma^{\mu}=2\eta^{\mu\nu}\mathbf{1}.
\end{equation}

Inserting $\bar{\sigma}^{\nu}\partial_{\nu}^{(\alpha)}$ in the eq.
\eqref{eq:Fractional Weyl 1}, we obtain 
\begin{equation}
i\left(\hbar\right)^{\alpha}\bar{\sigma}^{\nu}\partial_{\nu}^{(\alpha)}\sigma^{\mu}\partial_{\mu}^{(\alpha)}\chi_{L,\alpha}=0,
\end{equation}
 or symmetrizing 

The last equation can be rewritten with the help of the $\sigma^{\mu}$
properties as $i\left(\hbar\right)^{\alpha}\eta^{\mu\nu}\partial_{\nu}^{(\alpha)}\partial_{\mu}^{(\alpha)}\chi_{L,\alpha}=0$
that leads to the propagating fractional wave equation for left helicity
fermion $i\left(\hbar\right)^{\alpha}\square^{(\alpha)}\chi_{L,\alpha}=0.$
The notation for the box symbol it is not to be confused with the
fractional power operator in distribution theory. Here the the box
symbol is defined as $\square^{(\alpha)}\equiv\frac{M_{t,\alpha}}{c^{\alpha}}\frac{\partial^{\alpha}}{\partial t^{\alpha}}\frac{\partial^{\alpha}}{\partial t^{\alpha}}-M_{x,\alpha}^{2}\frac{\partial^{\alpha}}{\partial x^{\alpha}}\frac{\partial^{\alpha}}{\partial x^{\alpha}}.$

Right Helicity:

A similar procedure can be used to construct a projection for right
helicity spin, using in eq.\eqref{eq:projection of spin onto linear momentum}
$\lambda_{\alpha}=-1.$

Now the spinor field is noted as $\xi_{R}$ and belongs to a group
representation such as $\xi_{R}\in(0;1/2).$ Following the approach
sequence, the second fractional Weyl equation reads

\begin{equation}
i\left(\hbar\right)^{\alpha}\bar{\sigma}^{\mu}\partial_{\mu}^{(\alpha)}\xi_{R,\alpha}=0,
\end{equation}
 that also conduct us to the equation 
\begin{equation}
i\left(\hbar\right)^{\alpha}\square^{(\alpha)}\xi_{R,\alpha}=0.
\end{equation}

In compact notation, the fractional derivative is given by$\partial_{\mu}^{\alpha}=(\frac{1}{c^{\alpha}}\partial_{t}^{\alpha};M_{x,\alpha}\nabla^{\alpha})$

The two Weyl equations can be written in a more compact form in terms
of a four dimensional spinorial field $\Psi_{\alpha}$as

\begin{equation}
i\hbar^{\alpha}\left(\begin{array}{cc}
\mathbf{0} & \bar{\sigma}^{\mu}\partial_{\mu}^{\alpha}\\
\sigma^{\mu}\partial_{\mu}^{\alpha} & \mathbf{0}
\end{array}\right)\binom{\chi_{L,\alpha}}{\xi_{R,\alpha}}=i\hbar^{\alpha}\gamma^{\mu}\partial_{\mu}^{\alpha}\Psi_{\alpha}=0,
\end{equation}
where $\gamma^{\mu}$are usual the Dirac gamma matrices.

We proceed now by introducing a mass parameter that mixes the two
quiral components to obtain an equations capable to describe the dynamics
of a massive particle. To this intention we write the two fractional
Weyl equations as

\begin{equation}
\begin{cases}
i\left(\hbar\right)^{\alpha}\sigma^{\mu}\partial_{\mu}^{(\alpha)}\chi_{L,\alpha} & +u^{\alpha}c^{\alpha}\xi_{R,\alpha}=0\\
i\left(\hbar\right)^{\alpha}\bar{\sigma}^{\mu}\partial_{\mu}^{(\alpha)}\xi_{R,\alpha} & +\tilde{u}^{\alpha}c^{\alpha}\chi_{L,\alpha}=0
\end{cases},\label{eq:Weyl with mass term}
\end{equation}
where $u^{\alpha}$ and $\tilde{u}^{\alpha}$ are parameters which,
as we shall show below, will be the mass of the charged fermions.
Imposing that the above equations be compatible with the fractional
energy-momentum relation \cite{Aspects-nosso}

\begin{equation}
E^{2\alpha}=p^{2\alpha}c^{2\alpha}+m^{2\alpha}c^{4\alpha}.
\end{equation}

From the second of eq.\eqref{eq:Weyl with mass term}, we obtain 

\begin{equation}
\chi_{L,\alpha}=-\frac{i\hbar^{\alpha}}{\tilde{u}^{\alpha}c^{\alpha}}\bar{\sigma}^{\mu}\partial_{\mu}^{(\alpha)}\xi_{R,\alpha},
\end{equation}
that, substituted in the first of those equations and rewriting, yelds 

\begin{equation}
\hbar^{2\alpha}\sigma^{\mu}\partial_{\mu}^{(\alpha)}\bar{\sigma}^{\nu}\partial_{\nu}^{(\alpha)}\xi_{R,\alpha}+u^{\alpha}\tilde{u}^{\alpha}c^{2\alpha}\xi_{R,\alpha}=0,
\end{equation}
or 
\begin{equation}
\square^{(\alpha)}\xi_{R,\alpha}+\frac{u^{\alpha}\tilde{u}^{\alpha}c^{2\alpha}}{\hbar^{2\alpha}}\xi_{R,\alpha}=0.
\end{equation}

The above equation indicates by comparing with a fractional Klein
Gordon \cite{Aspects-nosso} equation that $u^{\alpha}=\tilde{u}^{\alpha}=-m^{\alpha}.$With
theses observations, the fractional Dirac equation can now be written
from \eqref{eq:Weyl with mass term} in the general case $\alpha\neq1$
as 

\begin{equation}
(i\hbar^{\alpha}\gamma^{\mu}\partial_{\mu}^{\alpha}-{\bf 1}m^{\alpha}c^{\alpha})\Psi_{\alpha}=0\label{eq:Fractional Dirac}
\end{equation}

\subsection{Minimal Coupling, Field Transformation and the Covariant Fractional
Derivative}

The conjugated Dirac equation is

\begin{equation}
\overline{\Psi}_{\alpha}(-i\hbar^{\alpha}\gamma^{\mu}\partial_{\mu}^{\alpha}-{\bf 1}m^{\alpha}c^{\alpha})=0.\label{eq:Conjug Fractional Dirac}
\end{equation}

Now, multiplying eq. \eqref{eq:Fractional Dirac} by $\overline{\Psi}$and
eq. \eqref{eq:Conjug Fractional Dirac} by $\Psi$ and subtracting
we obtain

\begin{equation}
\partial_{\mu}^{\alpha}\overline{\Psi}_{\alpha}\Psi_{\alpha}\gamma^{\mu}+\overline{\Psi}_{\alpha}\gamma^{\mu}\partial_{\mu}^{\alpha}\Psi_{\alpha}=\partial_{\mu}^{\alpha}(\overline{\Psi}_{\alpha}\gamma^{\mu}\Psi_{\alpha})=0.\label{eq:Fractional Continuity}
\end{equation}

The quantity in the bracket can be identified as a conserved current
and the equation is a fractional continuity equation. From standard
electromagnetism, we now that $j_{\mu}A^{\mu}$ is the electric current
coupled to electromagnetic field by a potential tensor $A^{\mu}$
, where $j_{\mu}$is of form $j_{\mu}=e\overline{\Psi}\gamma^{\mu}\Psi.$

For the fractional case, we can think of a minimal coupling term as
$e_{\alpha}\overline{\Psi}_{\alpha}\gamma^{\mu}\Psi_{\alpha}A_{\mu}^{\alpha}$,
where $A_{\mu}^{\alpha}=(\phi^{\alpha};-\vec{A}^{\alpha})$ is the
fractional potential tensor transform under Gauge transform as $(A_{\mu}^{\alpha})^{'}=A_{\mu}^{\alpha}+\partial_{\mu}^{\alpha}\chi$

Following the integer model of Lagrangian, we can write a fractional
Lagrangian with electromagnetic field coupled as 
\begin{equation}
\mathcal{L=}\overline{\Psi}_{\alpha}(i\hbar^{\alpha}\gamma^{\mu}\partial_{\mu}^{\alpha}-{\bf 1}m^{\alpha}c^{\alpha})\Psi_{\alpha}-e_{\alpha}\overline{\Psi}_{\alpha}\gamma^{\mu}\Psi_{\alpha}A_{\mu}^{\alpha},
\end{equation}
that can be written as 
\begin{equation}
\mathcal{L=}\overline{\Psi}_{\alpha}i\hbar^{\alpha}\gamma^{\mu}(\partial_{\mu}^{\alpha}+\frac{ie_{\alpha}}{\hbar^{\alpha}}A_{\mu}^{\alpha})\Psi_{\alpha}-{\bf 1}m^{\alpha}c^{\alpha}\overline{\Psi}_{\alpha}\Psi_{\alpha}.
\end{equation}

In order to the theory remain covariant, we assume a spinor field
transformation of form

\begin{equation}
\Psi_{\alpha}^{'}=R(\chi)\Psi_{\alpha},
\end{equation}
 where $R(\chi)$ have the unitary property and its explicit form
will be determined in the sequence. The Lagrangian density with the
electromagnetic coupling term lead us to define a fractional covariant
derivative of form

\begin{equation}
D_{\mu}^{\alpha}=\partial_{\mu}^{\alpha}+\frac{ie_{\alpha}}{c^{\alpha}\hbar^{\alpha}}kA_{\mu}^{\alpha},
\end{equation}
where the tensor field $A_{\mu}^{\alpha}$ is considered to have a
fractional Gage symmetry and transforms by

\begin{equation}
(A_{\mu}^{\alpha})^{'}=A_{\mu}^{\alpha}+\partial_{\mu}^{\alpha}\chi.
\end{equation}

The fractional covariant derivative of the field $\Psi_{\alpha}$
will have to obey the same field transform as the fields, or

\begin{equation}
(D_{\mu}^{\alpha}\Psi_{\alpha})^{'}=R(\chi)D^{\alpha}\Psi_{\alpha},
\end{equation}
In more details we can write

\begin{equation}
\partial_{\mu}^{\alpha}\Psi_{\alpha}^{'}+\frac{ie_{\alpha}}{c^{\alpha}\hbar^{\alpha}}kA_{\mu}^{\alpha'}\Psi_{\alpha}^{'}=R(\chi)\partial_{\mu}^{\alpha}\Psi_{\alpha}+\frac{ie_{\alpha}}{c^{\alpha}\hbar^{\alpha}}kA_{\mu}^{\alpha}\Psi_{\alpha}
\end{equation}
which results in a fractional differential equation of form

\begin{equation}
\frac{\partial_{\mu}^{\alpha}R}{R}=-\frac{ie_{\alpha}}{c^{\alpha}\hbar^{\alpha}}\partial_{\mu}^{\alpha}\chi.
\end{equation}

The solution of the above equation is

\begin{equation}
R=\exp(-\frac{ie_{\alpha}}{c^{\alpha}\hbar^{\alpha}}\chi),
\end{equation}
 This can be easily proven as follows. Fractionally deriving the above
equation, with the use of eq.\eqref{eq:chain rule space-time coarse-1}
results in

\begin{equation}
\partial_{\mu}^{\alpha}R=-\frac{ie_{\alpha}}{c^{\alpha}\hbar^{\alpha}}\exp(-\frac{ie_{\alpha}}{c^{\alpha}\hbar^{\alpha}}\chi)\partial_{\mu}^{\alpha}\chi=-\frac{ie_{\alpha}}{c^{\alpha}\hbar^{\alpha}}R\partial_{\mu}^{\alpha}\chi,
\end{equation}
which proves the assertion.

The fractional Dirac equation, in a coarse-grained scenario may now
be written with the minimal coupling as 
\begin{equation}
i\hbar^{\alpha}\gamma^{\mu}(\partial_{\mu}^{\alpha}+\frac{ie_{\alpha}}{c^{\alpha}\hbar^{\alpha}}A_{\mu}^{\alpha})\Psi_{\alpha}-{\bf 1}m^{\alpha}c^{\alpha}\Psi_{\alpha}=0
\end{equation}

Separating spacial and time terms, multiplying by $\gamma^{0}$ and
using the properties of gamma Dirac matrices, we obtain

\begin{equation}
i\hbar^{\alpha}\gamma^{0}\gamma^{0}\frac{1}{c^{\alpha}}\partial_{t}^{\alpha}\Psi_{\alpha}+i\hbar^{\alpha}\gamma^{0}\gamma^{i}M_{x,\alpha}\partial_{x}^{\alpha}\Psi_{\alpha}+i\hbar^{\alpha}\gamma^{0}\gamma^{0}\frac{ie_{\alpha}}{c^{\alpha}\hbar^{\alpha}}\phi^{\alpha}\Psi_{\alpha}+i\hbar^{\alpha}\gamma^{0}\gamma^{i}\frac{ie_{\alpha}}{c^{\alpha}\hbar^{\alpha}}(-\overrightarrow{A^{\alpha}})\Psi_{\alpha}-{\bf \gamma^{0}}m^{\alpha}c^{\alpha}\Psi_{\alpha}=0
\end{equation}
or

\begin{eqnarray}
i\hbar^{\alpha}{\bf 1}\partial_{t}^{\alpha}\Psi_{\alpha} & = & c^{\alpha}\gamma^{0}\gamma^{i}(-i\hbar^{\alpha}M_{x,\alpha}\partial_{x}^{\alpha}\Psi_{\alpha}-\frac{e_{\alpha}}{c^{\alpha}}\vec{A}^{\alpha}\Psi_{\alpha})+\gamma^{0}\gamma^{0}e_{\alpha}\phi^{\alpha}\Psi_{\alpha}+\gamma^{0}m^{\alpha}c^{2\alpha}\Psi_{\alpha}\\
i\hbar^{\alpha}{\bf 1}\partial_{t}^{\alpha}\Psi_{\alpha} & = & c^{\alpha}\gamma^{0}\gamma^{i}(-i\hbar^{\alpha}M_{x,\alpha}\partial_{x}^{\alpha}\Psi_{\alpha}-\frac{e_{\alpha}}{c^{\alpha}}\overrightarrow{A^{\alpha}}\Psi_{\alpha})+\mathbf{1}e_{\alpha}\phi^{\alpha}\Psi_{\alpha}+\gamma^{0}m^{\alpha}c^{2\alpha}\Psi_{\alpha}
\end{eqnarray}

Using the correspondence principle, we have

\begin{equation}
i\hbar^{\alpha}{\bf 1}\partial_{t}^{\alpha}\Psi_{\alpha}=c^{\alpha}\overrightarrow{\alpha}\circ(\overrightarrow{p^{\alpha}}-\frac{e_{\alpha}}{c^{\alpha}}\overrightarrow{A^{\alpha}})\Psi_{\alpha}+\hbar^{\alpha}\mathbf{1}e_{\alpha}\phi^{\alpha}\Psi_{\alpha}+\gamma^{0}m^{\alpha}c^{2\alpha}\Psi_{\alpha}.\label{eq:Dirac com acopla min}
\end{equation}

Now we write the Dirac spinor as $4\times1$ matrix in a coarse-grained
scenario as

\begin{equation}
\Psi_{\alpha}=\binom{\psi_{\alpha,s}}{\psi_{\alpha,w}},
\end{equation}
where we have named the strong($s$) and weak($w$) components as
a $2\times1$ matrices as $\psi_{\alpha,s}=\binom{\psi_{\alpha,1}}{\psi_{\alpha,2}},$
$\psi_{\alpha,w}=\binom{\psi_{\alpha,3}}{\psi_{\alpha,4}},$ respectively.
In a sympletic form we can write

\begin{equation}
\begin{cases}
\psi_{\alpha,s} & =\psi_{\alpha,1}+i\psi_{\alpha,2}\\
\psi_{\alpha,w} & =\psi_{\alpha,3}-i\psi_{\alpha,4}
\end{cases}.
\end{equation}

\subsection{Non-Relativistic Limit of Fractional Dirac Equation and the Fractional
g-factor}

In order to proceed with the non-relativistic limit of the fractional
Dirac Equation, we have to consider the dominant term in the Hamiltonian
as the rest energy given by $m^{\alpha}c^{2\alpha}.$ We then propose
an ansatz for the solution to the fractional Dirac Equation as 
\begin{equation}
\Psi_{\alpha}^{'}=e^{(-i\frac{E^{\alpha}}{\hbar^{\alpha}}t^{\alpha})}\binom{\psi_{\alpha,s}}{\psi_{\alpha,w}},\; E^{\alpha}\cong m^{\alpha}c^{2\alpha}.
\end{equation}

Inserting this ansatz into the fractional Dirac equation, eq.\eqref{eq:Fractional Dirac},
with the help of the chain rule \eqref{eq:Chainrule nondif func-1}
and the Leibniz rule \eqref{eq:Leibniz}, we obtain

\begin{equation}
i\hbar^{\alpha}{\bf 1}\partial_{t}^{\alpha}\binom{\psi_{\alpha,s}}{\psi_{\alpha,w}}=\binom{c^{\alpha}\overrightarrow{\sigma}\circ(\overrightarrow{p^{\alpha}}-\frac{e_{\alpha}}{c^{\alpha}}\overrightarrow{A^{\alpha}})}{c^{\alpha}\overrightarrow{\sigma}\circ(\overrightarrow{p^{\alpha}}-\frac{e_{\alpha}}{c^{\alpha}}\overrightarrow{A^{\alpha}})}\binom{\psi_{\alpha,w}}{\psi_{\alpha,s}}+e_{\alpha}\phi^{\alpha}\binom{\psi_{\alpha,s}}{\psi_{\alpha,w}}+m^{\alpha}c^{2\alpha}\binom{\psi_{\alpha,s}}{-\psi_{\alpha,w}}-m^{\alpha}c^{2\alpha}\Gamma(\alpha+1)\binom{\psi_{\alpha,s}}{\psi_{\alpha,w}},
\end{equation}
that furnish two equations as follows 
\begin{equation}
\begin{cases}
i\hbar^{\alpha}{\bf 1}\partial_{t}^{\alpha}\psi_{\alpha,b} & =c^{\alpha}\overrightarrow{\sigma}\circ(\overrightarrow{p^{\alpha}}-\frac{e_{\alpha}}{c^{\alpha}}\overrightarrow{A^{\alpha}})\psi_{\alpha,w}+e_{\alpha}\phi^{\alpha}\psi_{\alpha,s}+(1-\Gamma(\alpha+1))m^{\alpha}c^{2\alpha}\psi_{\alpha,s}\\
i\hbar^{\alpha}{\bf 1}\partial_{t}^{\alpha}\psi_{\alpha,w} & =c^{\alpha}\overrightarrow{\sigma}\circ(\overrightarrow{p^{\alpha}}-\frac{e_{\alpha}}{c^{\alpha}}\overrightarrow{A^{\alpha}})\psi_{\alpha,s}+e_{\alpha}\phi^{\alpha}\psi_{\alpha,w}-(1+\Gamma(\alpha+1))m^{\alpha}c^{2\alpha}\psi_{\alpha,w}
\end{cases}.\label{eq:Strong-Weak system}
\end{equation}
Considering now that the mass terms is dominant over the electrostatic
one, that is, $e_{\alpha}\phi^{\alpha}\lll m^{\alpha}c^{2\alpha}(1+\Gamma(\alpha+1))$,
and that the weak component fields $\psi_{\alpha,w}$ has slow evolution,
when compared to the rest energy, $i\hbar^{\alpha}{\bf 1}\partial_{t}^{\alpha}\psi_{\alpha,w}<m^{\alpha}c^{2\alpha}(1+\Gamma(\alpha+1))\psi_{\alpha,w}.$
With these approximations, we can write for the second equation 
\begin{equation}
0\cong c^{\alpha}\overrightarrow{\sigma}\circ(\overrightarrow{p^{\alpha}}-\frac{e_{\alpha}}{c^{\alpha}}\overrightarrow{A^{\alpha}})\psi_{\alpha,s}+e_{\alpha}\phi^{\alpha}\psi_{\alpha,w}-(1+\Gamma(\alpha+1))m^{\alpha}c^{2\alpha}\psi_{\alpha,w},
\end{equation}
that leads to the relations between weak and strong field components
as 
\begin{equation}
\psi_{\alpha,w}\cong\frac{c^{\alpha}\overrightarrow{\sigma}\circ(\overrightarrow{p^{\alpha}}-\frac{e_{\alpha}}{c^{\alpha}}\overrightarrow{A^{\alpha}})}{(1+\Gamma(\alpha+1))m^{\alpha}c^{2\alpha}}\psi_{\alpha,s}.
\end{equation}

The above expression gives immediately that the weak component is,
in the scope of adopted approximations, very lower that the strong
one, $\psi_{\alpha,w}\lll\psi_{\alpha,s}.$

Now inserting this result into the first equation \eqref{eq:Strong-Weak system},
results that one equation for the strong component written as 
\begin{equation}
i\hbar^{\alpha}{\bf 1}\partial_{t}^{\alpha}\psi_{\alpha,s}=\frac{\overrightarrow{\sigma}\circ(\overrightarrow{p^{\alpha}}-\frac{e_{\alpha}}{c^{\alpha}}\overrightarrow{A^{\alpha}})\overrightarrow{\sigma}\circ(\overrightarrow{p^{\alpha}}-\frac{e_{\alpha}}{c^{\alpha}}\overrightarrow{A^{\alpha}})}{(1+\Gamma(\alpha+1))m^{\alpha}}\psi_{\alpha,s}+e_{\alpha}\phi^{\alpha}\psi_{\alpha,s}+(1-\Gamma(\alpha+1))m^{\alpha}c^{2\alpha}\psi_{\alpha,s}.
\end{equation}

Defining the fractional momentum operator as $\overrightarrow{\pi^{\alpha}}\equiv\overrightarrow{p^{\alpha}}-\frac{e_{\alpha}}{c^{\alpha}}\overrightarrow{A^{\alpha}},$
and using the well known propriety of Pauli matrices $\overrightarrow{(\sigma}\circ\pi^{\alpha})\overrightarrow{(\sigma}\circ\pi^{\alpha})=\pi^{\alpha}\circ\pi^{\alpha}+i\overrightarrow{\sigma}\circ(\pi^{\alpha}\wedge\pi^{\alpha})$
and also that $(\pi^{\alpha}\wedge\pi^{\alpha})=-\frac{e_{\alpha}}{c^{\alpha}}\overrightarrow{p^{\alpha}}\wedge\overrightarrow{A^{\alpha}}.$
Now using the definition $\overrightarrow{p^{\alpha}}=i\hbar^{\alpha}M_{x,\alpha}\vec{\nabla}^{\alpha},$$\vec{\nabla}^{\alpha}\equiv\hat{i}\frac{\partial^{\alpha}}{\partial x^{\alpha}}+\hat{j}\frac{\partial^{\alpha}}{\partial y^{\alpha}}+\hat{k}\frac{\partial^{\alpha}}{\partial z^{\alpha}}$
and that in analogy with integer case, $\vec{\nabla}^{\alpha}\wedge\overrightarrow{A^{\alpha}}=\overrightarrow{B^{\alpha}},$
than 
\begin{equation}
i\hbar^{\alpha}{\bf 1}\partial_{t}^{\alpha}\psi_{\alpha,b}=\frac{(\overrightarrow{p^{\alpha}}-\frac{e_{\alpha}}{c^{\alpha}}\overrightarrow{A^{\alpha}})^{2}}{(1+\Gamma(\alpha+1))m^{\alpha}}\psi_{\alpha,b}-\frac{\hbar^{\alpha}M_{x,\alpha}\frac{e_{\alpha}}{c^{\alpha}}\overrightarrow{\sigma}\circ\overrightarrow{B^{\alpha}}}{(1+\Gamma(\alpha+1))m^{\alpha}}\psi_{\alpha,b}+e_{\alpha}\phi^{\alpha}\psi_{\alpha,b}+(1-\Gamma(\alpha+1))m^{\alpha}c^{2\alpha}\psi_{\alpha,b}.\label{eq:Fractional Pauli}
\end{equation}

The eq. \eqref{eq:Fractional Pauli} is the fractional version of
the Pauli equation in a coarse-grained scenario.

Remembering that the spin term is $\overrightarrow{S_{\alpha}}=\frac{\hbar^{\alpha}}{2}\Gamma(\alpha+1)M_{x,\alpha}\overrightarrow{\sigma}$,
the second term in the rhs of the above equation can be written as
\begin{equation}
-\frac{e_{\alpha}}{2m^{\alpha}c^{\alpha}}\left(\frac{4}{(1+\Gamma(\alpha+1))\Gamma(\alpha+1)}\right)\overrightarrow{S_{\alpha}}\circ\overrightarrow{B^{\alpha}}\psi_{\alpha,s}.
\end{equation}

The factor into the bracket can be identified as the fractional g-factor
$g_{frac},$
\begin{equation}
g_{frac}=\frac{4}{(1+\Gamma(\alpha+1))\Gamma(\alpha+1)}.\label{eq:g-frac}
\end{equation}

Note the when $\alpha=1,$ $g_{frac}=g=2$ and the equation becomes
the usual Pauli equation.

We can map a fractional parameter with the CODATA \cite{CODATA} known
value of $g_{exp}$ for electrons and muons and also ref.\cite{Passera}
for taus particles. The mapping can be done by solving numerically
the equation 
\begin{equation}
g_{frac}=\frac{4}{(1+\Gamma(\alpha+1))\Gamma(\alpha+1)}=g_{exp}.\label{eq:g-frac comparative teo-exp}
\end{equation}

In our Discussion and Conclusions (Section 6), we shall discuss in
more details this result and how we make use of it to fit the $g-factors$
of the charged leptonic particles.

\section{The Gordon Decomposition for Fractional Dirac Equation:The Fractional
Spin Current}

From eq.\eqref{eq:Fractional Continuity} we can define the fractional
current density as 
\begin{equation}
j_{\alpha}^{\mu}=\overline{\Psi}_{\alpha}\gamma^{\mu}\Psi_{\alpha}.
\end{equation}

The above equation may be rewritten as 

\begin{equation}
j_{\alpha}^{\mu}=\frac{1}{2}\left(\overline{\Psi}_{\alpha}\gamma^{\mu}\Psi_{\alpha}+\overline{\Psi}_{\alpha}\gamma^{\mu}\Psi_{\alpha}\right).\label{eq:Current Symmetric}
\end{equation}

From the Dirac equation eq. \eqref{eq:Fractional Dirac} and its conjugated
\eqref{eq:Conjug Fractional Dirac} we may write, respectively, 
\begin{eqnarray}
\Psi_{\alpha} & = & \frac{i\hbar^{\alpha}}{m^{\alpha}c^{\alpha}}\gamma^{\mu}\partial_{\mu}^{\alpha}\Psi_{\alpha},\\
\overline{\Psi}_{\alpha} & = & -\frac{i\hbar^{\alpha}}{m^{\alpha}c^{\alpha}}\partial_{\mu}^{\alpha}\overline{\Psi}_{\alpha}\gamma^{\mu}.
\end{eqnarray}

Now, inserting these results in the first and second terms of the
eq. \eqref{eq:Current Symmetric}, respectively, we can write for
the fractional current 
\begin{equation}
j_{\alpha}^{\mu}=\frac{i\hbar^{\alpha}}{m^{\alpha}c^{\alpha}}\left(\left(\overline{\Psi}_{\alpha}\gamma^{\mu}\gamma^{\nu}\right)\left(\partial_{\nu}^{\alpha}\Psi_{\alpha}\right)-\left(\partial_{\nu}^{\alpha}\overline{\Psi}_{\alpha}\right)\left(\gamma^{\nu}\gamma^{\mu}\Psi_{\alpha}\right)\right).\label{eq:Current extended}
\end{equation}

For the gamma matrix we can write that

\begin{eqnarray}
\gamma^{\mu}\gamma^{\nu} & = & \frac{1}{2}\left(\gamma^{\mu}\gamma^{\nu}+\gamma^{\nu}\gamma^{\mu}\right)+\frac{1}{2}\left(\gamma^{\mu}\gamma^{\nu}-\gamma^{\nu}\gamma^{\mu}\right)\nonumber \\
 & = & \eta^{\mu\nu}1+\Sigma^{\mu\nu},
\end{eqnarray}

and
\begin{eqnarray}
\gamma^{\nu}\gamma^{\mu} & = & \eta^{\mu\nu}1+\Sigma^{\nu\mu}=\eta^{\mu\nu}1-\Sigma^{\mu\nu}
\end{eqnarray}

where $\Sigma^{\mu\nu}=\frac{1}{2}\left(\gamma^{\mu}\gamma^{\nu}-\gamma^{\nu}\gamma^{\mu}\right)=\frac{1}{2}\left[\gamma^{\mu},\gamma^{\nu}\right]$
and $\eta^{\mu\nu}$ is the metric tensor.

Using the above definitions and inserting into the eq. \eqref{eq:Current extended},
we obtain for the current a more decoupled form 
\begin{eqnarray}
j_{\alpha}^{\mu} & = & \frac{i\hbar^{\alpha}}{2m^{\alpha}c^{\alpha}}\left(\overline{\Psi}_{\alpha}\left(\partial_{\nu}^{\alpha}\Psi_{\alpha}\right)-\left(\partial_{\nu}^{\alpha}\overline{\Psi}_{\alpha}\right)\Psi_{\alpha}\right)+\nonumber \\
 & + & \frac{i\hbar^{\alpha}}{2m^{\alpha}c^{\alpha}}\left(\overline{\Psi}_{\alpha}\Sigma^{\mu\nu}\left(\partial_{\nu}^{\alpha}\Psi_{\alpha}\right)+\left(\partial_{\nu}^{\alpha}\overline{\Psi}_{\alpha}\right)\Sigma^{\mu\nu}\Psi_{\alpha}\right).
\end{eqnarray}

Defining $\sigma^{\mu\nu}=i\Sigma^{\mu\nu},$ the second term in the
above equation may be identified with the spin contribution to the
fractional current and reads as 
\begin{equation}
j_{\alpha,spin}^{\mu}=\frac{\hbar^{\alpha}}{2m^{\alpha}c^{\alpha}}\partial_{\nu}^{\alpha}\left(\overline{\Psi}_{\alpha}\sigma^{\mu\nu}\Psi_{\alpha}\right).
\end{equation}

Note that the above decomposition is performed in the configuration
space instead of the Fourier space, because we are not dealing with
generalized functions, but with non differentiable functions.

\bigskip{}

\section{Discussion and Conclusions}

With the help of the eq. \eqref{eq:g-frac comparative teo-exp} we
can compare the the fractional parameter $\alpha$ for electrons,
muons and taus. The numerical results show that the fractionalities,
$\alpha$, exhibit a hierarchy,$\alpha(electron)>\alpha(muon)>\alpha(tau)$:

For the electron (CODATA):$g_{e-Exp}=2.00231930436146$, $\alpha_{e-Exp}=0.9981697906061296726$,
indicating a system with low fractionality ($\alpha$ near 1).

For the muon (CODATA):$g_{\mu-Exp}=2.00233184182$, $\alpha_{\mu-Exp}=0.9981598882409105161$.

For the tau (QED, see eg. \cite{Passera}): $g_{\tau}=2.00235442$,
$\alpha_{\tau}=0.998142055249517567.$

The higher is the $g-factor$, the farther is the $alpha$-parameter
from $\alpha=1$ (g-fator is a decreasing function with alpha, as
can be seen numerically from eq. \eqref{eq:g-frac}). Note that a
bigger $alpha$ means fractionality closer to $1$. Moreover, if this
parameter deviates from $\alpha=1,$ it may indicate that the system
is more sparse (roughness is larger), ie, there is a smaller range
about the different types of interactions (particles and fields) in
the vicinity of pseudo particule. Therefore, if the particle interacts
less or have a lower mean life than other, the $alpha$-parameter
would be a little more distant from one (in the downward direction),
indicating a lower complexity of interactions. The closer $\alpha$
is to $1$ indicates a greater complexity of these interactions (a
particle interacting more, more structures can interact around). Alpha
equal to $1$ means one type of total mixing memory (providing an
idealized integer model, which is not natural), with the pseudo partícules
feeling the interactions of all kinds, so that, on average, these
fluctuations are nullify (a statistical average).

In terms of mean life, muon has $10^{-6}s$, tau$10^{-15}s$, while
electrons are stable. The relation with the $alpha$-parameter can
be understood as follows: The lower is the mean life means that the
particle does not have sufficient time to interact with the surrounding
environment and with other particles, implying that the $alpha$-parameter
is more distant from 1. The interaction universe seen by those particles
is limited to the closer interacting particles. Stable particles like
electrons can ``see'' the entire environment and interact more,
justifying a fractional parameter closer to 1 than the other particles.
The time scale of the interaction evidenced by the mean life can gives
clues of the fractionality.

From QED calculations\cite{CODATA,Eduardo de Rafael 2012}, we also
have for electron, muon and tau:

\textbf{$g_{e-QED}=2.00231930436364,$ $\alpha_{e-QED}=0.9981697906044078681;$}

\textbf{$g_{\mu-QED}=2.002331694362,$$\alpha_{\mu-QED}=0.9981600047070900225$.}

Comparing the parameters above with those from the measured anomalies\cite{CODATA},
we can see that\textbf{ $g_{e-QED}=2.00231930436364>g_{e-Exp}=2.00231930436146$}
and leads to\textbf{ $\alpha_{e-QED}=0.9981697906044078681<\alpha_{e-Exp}=0.9981697906061296726$
.}

and also that

\textbf{$g_{\mu-QED}=2.002331694362<g_{\mu-Exp}=2.00233184182$ }and
leads to\textbf{$\alpha_{\mu-QED}=0.9981600047070900225>\alpha_{\mu-Exp}=0.9981598882409105161.$}

For the electron, the results may be indicating that the complexity
of the interactions taken into account in the QED calculation may
be lower than it would be in the in the reality seen by the experiment.
That is, the electron, as a pseudo-particle, keeps hidden other interactions
that are not well described or are incomplete in the SM description.
Thus, the results are different for QED compared to the experimental
results. In the experimental reality, the complexiy is greater than
that considered in the QED calculations based on SM. That is, the
fractionality should indicate that the SM, although very good, may
not be providing all information necessary to describe the interactions
in a more complete view, either at high energies or in granular or
fractal space-time. Thus, the FC may gives evidences that the SM could
require corrections (or higher order calculations by QED).

For the muon, an opposite behaviour in terms of fractional parameter
could be observed and indicated that the QED calcultions might be
taken into account more interactions and consequnt complexity than
the particle realy experiment and, again, signaling that the SM may
not be complete to describe the whole interaction scenario.

In summary, in the present work we\textbf{ }have built up a fractional
Dirac equation in a coarse-grained scenario by taking into account
a fractional Weyl equation, a fractional angular momentum algebra,
introducing a mass parameter and imposing that the equations be compatible
with the fractional energy-momentum relation. Considering then a minimal
coupling in the Lagrangian and introducing a field transformation
to reveal a covariant fractional derivative, the free Lagrangian density
with the electromagnetic coupling term leads us to define a fractional
covariant derivative. In the sequel, we proceeded by a non-relativistic
limit of the Dirac equation to obtain a fractional version of the
Pauli equation. We than have investigated the anomalous magnetic moment
for the charged leptons, electrons, muons and taus. By the a fractional
approach we have obtained, In each the cases of study, that the results
agrees with standard integer order in the convenient limits. 

We have shown that a mapping of the anomalous magnetic g factor is
possible in terms of a fractional parameter.

Defining a fractional current density we have also performed a fractional
Gordon decomposition and identified the spin contribution to the fractional
current density.

We suggested that the understanding of the results comparatively for
electrons, muons e taus may be thought in the realm of complexity,
mean life and pseudo-particles concepts and we had shown that $\alpha(electron)>\alpha(muon)>\alpha(tau)$.
We established a connection between mean life of the particle and
fractionality, showing that small mean life can leads to a fractional
parameter more distant from $\alpha=1$ than for particles with more
stability and consequently greater mean life than the other one.

The complexity of the interactions involved may hide the detailed
knowledge on its dynamical aspects, since the non-local characteristics
of the interactions. Therefore, a detailed description may be very
difficult to be obtained or even be impossible. This can explain the
lacking of a precise formulation of theories. The theory may has beyond
the standard model, since we think that the interaction aspects due
complexity has to be taken into account. Therefore we suggest that
this can be done with an effective theories, capable to give some
indicative evidences for the existence of complex interactions. Here
we think that the fractional calculus may be a good candidate for
this effective theories.

\textbf{\bigskip{}
}

Cresus F. L. Godinho is acknowledged for the discussions at an early
stage of this work.

The authors wish to express their gratitude to FAPERJ-Rio de Janeiro
and CNPq-Brazil for the partial financial support.\textbf{\bigskip{}
}

\end{document}